\documentclass[%
reprint,
superscriptaddress,
 amsmath,amssymb,
aps,
prl,
]{revtex4-1}

\usepackage{graphicx}
\usepackage{dcolumn}
\usepackage{bm}

\begin{document}

\preprint{APS/123-QED}

\title{Neutron Brillouin scattering with pulsed spallation neutron source\\
- spin-wave excitations from ferromagnetic powder samples -
}

\author{S. Itoh}
\affiliation{Institute of Materials Structure Science, High Energy Accelerator Research Organization, Tsukuba 305-0801, Japan}
\author{Y. Endoh}
\affiliation{Institute of Materials Structure Science, High Energy Accelerator Research Organization, Tsukuba 305-0801, Japan}
\affiliation{Institute for Materials Research, Tohoku University, Sendai 980-8577, Japan}
\author{T. Yokoo}
\author{D. Kawana}
\affiliation{Institute of Materials Structure Science, High Energy Accelerator Research Organization, Tsukuba 305-0801, Japan}
\author{Y. Kaneko} 
\affiliation{Correlated Electron Research Group (CERG), RIKEN Advanced Science Institute, Wako, Saitama 351-0198, Japan} 
\author{Y.Tokura} 
\affiliation{Correlated Electron Research Group (CERG), RIKEN Advanced Science Institute, Wako, Saitama 351-0198, Japan} 
\affiliation{Department of Applied Physics, The University of Tokyo, Tokyo 113-8656, Japan} 
\author{M. Fujita}
\affiliation{Institute for Materials Research, Tohoku University, Sendai 980-8577, Japan}


\date{\today}

\begin{abstract}
Neutron Brillouin scattering (NBS) method was developed using a pulsed spallation neutron source, and the feasibility of NBS was demonstrated by observing ferromagnetic spin waves in La$_{0.8}$Sr$_{0.2}$MnO$_3$ and SrRuO$_3$ powders. Gapless spin-wave excitations were observed in La$_{0.8}$Sr$_{0.2}$MnO$_3$, which were continuously extended to the lower scattering vector $Q$ from previous results using single crystals. The novel result is a well-defined quadratic $Q$ dependence in the spin-wave dispersion curve with a large energy gap in SrRuO$_3$ indicating robust spin-orbit coupling.
\begin{description}
\item[PACS numbers] 75.30.Ds, 75.40.Gb, 75.50.Gg
\end{description}
\end{abstract}

\maketitle


In investigations of spin mediated materials physics, it is essential to experimentally determine microscopic interaction energies. For instance, an exchange interaction within the mean field approximation corresponds to the ordering temperature. This value is directly derived by the momentum dependence of spin waves in ordered states. Thus, spectroscopic investigations using thermal neutrons have been established as the most valuable and powerful experimental methods, although actual experiments have been performed mainly with single crystals, which require large volume and high quality samples from advanced spectrometers installed even at the most intense neutron sources.

The current research trend in material sciences requires new materials of high performance as well as energy effective functioning for device applications. To support this requirement for magnetic materials as one typical example, ferromagnetic semiconductors are a key material for future spintronics devices, and they have already been proven to be functional at low temperatures. However, now, research directs to the discovery of new materials of high performance at ambient temperatures. Obviously, these materials are required to possess a high ordering temperature (Curie temperature $T_c$ for ferromagnets). Therefore, we need to seek magnetic materials with a large spin-wave stiffness constant and a zero energy gap, which show that they are easily magnetized. Most of newly synthesized materials are chemically and structurally complicated such that large single crystals are difficult to be synthesized. Neutron spectroscopy experiments require the use of powder samples, and sometimes, low quality materials with impurity phases.

The neutron Brillouin scattering (NBS) method is the most suitable for measuring the momentum dependence of spin-wave excitations from (000) in the forward direction. Here, NBS is defined in comparison with the optical Brillouin Scattering. The principles of NBS are not new, although, until recently, such experiments have been limited to a very small energy range, because intense neutrons have been available in the thermal energy region. Sub-eV neutrons are now used high flux neutron sources. In this work, we demonstrate the first clear result of spin-wave excitations from two powder ferromagnets by using a pulsed spallation neutron source: the colossal magnetoresistance manganite La$_{0.8}$Sr$_{0.2}$MnO$_3$ ($T_c$ = 316 K) \cite{endoh97,moussa07} and SrRuO$_3$ ($T_c$ = 165 K) \cite{klein96}. Both perovskites essentially possess a cubic lattice structure with a tiny distortion. The former is a typical example of a next generation magnetic device. The latter has recently become attractive because of an enhanced anomalous Hall effect due to robust spin-orbit coupling of the Ru 4d orbital \cite{fang03}. Large single crystals are not always available and, in particular, SrRuO$_3$ has not been synthesized as yet. However, powder samples are readily available. Scattering intensities from powder samples rapidly decrease as the scattering vector $Q$ increases because of the powder average of the dynamical structure factor. Scattering intensities from ferromagnets are strong only in the forward direction near (000). Owing to the kinematic constraints of neutron spectroscopy, incident energy ($E_i$) in the sub-eV region is necessary to measure scattering in a meV transferred-energy ($\omega$) range, and scattered neutrons need to be detected at very small scattering angles ($\phi$) of $1^{\circ}$ or less, with a high energy resolution of $\Delta\omega/E_i$.

\begin{figure}
\includegraphics[width=7.5cm]{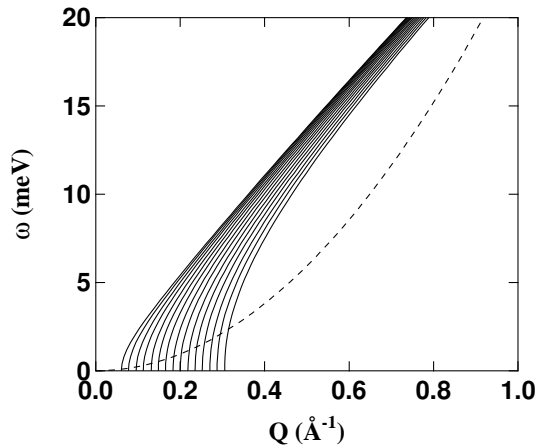}
\caption{\label{fig:epsart} Scan loci with $E_i$ = 102 meV at the centers of PSDs, which are located at the scattering angles from $0.5^{\circ}$ to $2.8^{\circ}$ (from left to right) in steps of $0.15^{\circ}$, used in the present experiment on the HRC. The $(Q,\omega)$ region above the dashed line (higher transfer energies) can never be accessed by a spectrometer with the lowest scattering angle of $\phi = 5^{\circ}$.}
\end{figure}

Efforts for realizing such experimental conditions in order to apply the NBS method have been made at the High Resolution Chopper Spectrometer (HRC) installed at the Material and Life Science Facility (MLF) in the Japan Proton Accelerator Research Complex (J-PARC) \cite{itoh11,yano11}.  Similarly, Pharos, at the pulsed spallation neutron source at the Los Alamos Neutron Science Center (LANSCE) in the Los Alamos National Laboratory \cite{robinson98} and BRISP at the steady state neutron source in the Institut Laue-Langevin (ILL) \cite{aisa06} have been applied NBS. On the HRC, polychromatic neutron beams extracted from a 25 Hz pulsed neutron source were monochromatized by a Fermi chopper. The energy of the monochromatic beams can be adjusted to the type of measurement. The neutrons scattered by the sample were then detected, and the energy transfer ($\omega$) was determined from the time-of-flight of the detected neutrons. For NBS, scattered neutrons were detected by a two-dimensional array of position sensitive detectors (PSDs) at a small $\phi$. In the array, linear PSDs were mounted with their linear axes vertical, and the centers of the PSDs were located horizontally at $\phi$ between $0.5^{\circ}$ and $2.8^{\circ}$ in constant steps of approximately $0.15^{\circ}$ \cite{yano11}. 
The horizontal divergence of the neutron beams was controlled by a collimator mounted immediately upstream of the sample. The collimator was composed of slits of vertical cadmium sheets with a collimation of $0.3^{\circ}$. 
The measurement was performed at the temperatures $T$ = 6 K and 254 K for La$_{0.8}$Sr$_{0.2}$MnO$_3$ and at $T$ = 7 K for SrRuO$_3$. Since an estimation of background contribution was important, an empty scan under exactly the same geometry as the sample scan, with only the sample crystal removed, was also performed.

The scan loci of the centers of the PSDs in the array on the HRC with $E_i$ = 102 meV selected in the present experiment on the $(Q,\omega)$ space are presented in Fig.1. A detector element located at $\phi$ scans points in the $(Q,\omega)$ space as described by
\begin{equation}
\frac{\hbar^2Q^2}{2m} = 2E_i - \omega - 2\left(E_i(E_i-\omega)\right)^{1/2} \cos \phi,
\end{equation}
where $\hbar$ and $m$ are Plank's constant divided by $2\pi$ and the neutron mass, respectively. It should be noted that low angle detectors are essential to access the present $(Q,\omega)$ space. In fact, the region above the dashed line in Fig. 1, which is $\omega = (\hbar^2Q^2/2m)/\sin\phi$, or the envelope of scan loci for $\phi = 5^{\circ}$ in eq. (1) with respect to $E_i$, can never be accessed by a spectrometer having the lowest scattering angle of $\phi = 5^{\circ}$, for instance. The high energy-resolution $\Delta\omega/E_i$, which was a few percent or less, was achieved with short-pulse neutrons extracted from a decoupled moderator (a neutron source for high resolution experiments) and delivered to the HRC. The $Q$ resolution can be derived as follows from the derivatives of $\omega$ and $\phi$ in eq. (1):  
\begin{equation}
\left(\frac{\Delta Q}{k_i}\right)^2 = \left(\frac{1}{2} \frac{\Delta\omega}{E_i} \sin \frac{\phi}{2}\right)^2 + \left(\Delta\phi \cos \frac{\phi}{2}\right)^2,
\end{equation}
where $\hbar^2k_i^2/2m = E_i$. The angular resolution for a small $\phi$ is given by
\begin{equation}
\left(\Delta\phi\right)^2 = \frac{8\ln2}{12} \left( \frac{w_m^2+h_m^2}{L_1^2} + \frac{w_s^2+h_s^2}{L'^ 2} + \frac{w_d^2+h_d^2}{L_2^2} \right),
\end{equation}
where the symbols $w$ and $h$ indicate the width and height, respectively, $m$, $s$, and $d$ denote the moderator (neutron source), sample and detector, respectively; and $L'^{-1} = L_1^{-1} + L_2^{-1}$, where $L_1$ = 15 m and $L_2$ = 5.2 m are the distances from neutron source to the sample, and from the sample to the detector, respectively. Therefore, we can calculate $\Delta Q$ with the following parameters: $w_m$ = 100 mm, $h_m = \alpha w_m$, $w_s$ = $h_s$ = 30 mm, and $w_d$ = 12.7 mm. The HRC faces the moderator area of 100 mm $\times$ 100 mm, and an intensity gain of a factor of 3 was obtained for $E_i$ = 102 meV by a guide tube. The collimator cut the intensity gain in the horizontal direction, because the collimation was almost identical to the beam divergence between the moderator and the sample without the guide tube. Therefore, $\alpha = 3^{1/2}$ was used. For the plot of $h_d$ = 92 mm in Fig. 2, the angular resolution was calculated to be $\Delta\phi$ = 17 mrad. In the present experiment, since the first term in eq. (2) was much smaller than the second term because of $\Delta\omega$ determined below, the $Q$ resolution could be approximated as $\Delta Q = k_i \Delta\phi = 0.12$ \AA$^{-1}$.  Figure 2 presents the energy spectra observed at each fixed $\phi$ on the scan locus determined in eq. (1) after subtracting the empty scan data, where the neutron intensities were summed up over a detector length of 92 mm ($\pm46$ mm from the center position). Each spectrum shown in Fig. 2 was fitted to the following spectral function:
\begin{equation}
I(\omega) = \frac{A\left(n(\omega)+1\right) F(Q)^2\omega\omega_0\Gamma^2} {\left(\omega^2-\omega_0^2\right)^2+\left(\omega\Gamma\right)^2} + B\exp\left(-4\ln2\frac{\omega^2}{\Delta \omega^2}\right),
\end{equation}
for $\omega > 0$ meV by parameterizing $A$, $B$, $\omega_0$, $\Gamma$, and $\Delta\omega$, where $n(\omega)+1 = [1-\exp(-\omega/k_BT)]^{-1}$ with $k_B$ being the Boltzmann constant. For $\omega < 0$ meV, the temperature factor $n(\omega)+1$ in eq. (4) was replaced by the product of the temperature factor and the detailed balance factor, $n(\omega)\exp(\omega/k_BT)$. The magnetic form factor is denoted by $F(Q)$ [10]. The first and second terms in eq. (4) correspond to a spin wave component and elastic scattering with the energy resolution width $\Delta\omega$, respectively. The spin wave component is thus presented by the parameters $A$, $\omega_0$ and $\Gamma$. The peak intensity parameter $A$ was equivalent for both $T$ = 6 and 245 K: $A = 0.28 \pm 0.07$ and $0.26 \pm 0.01$ in the units in Fig. 2 for $\phi = 0.64^{\circ}$, respectively, for instance, and therefore, the intensity difference between $T$ = 6 and 245 K can be explained by the temperature factor. The elastic component consists of a Bragg tail around (000) and incoherent elastic scattering. In fact, the width of the plot of $B$ as a function of $Q$, which was converted for an appropriate $\phi$ and with $\omega$ = 0 meV from eq. (1), was obtained at 0.11 \AA$^{-1}$ at the full width at half maximum by fitting to a Gaussian function centered at $Q$ = 0 \AA$^{-1}$ plus a constant. The observed width was consistent with the above-calculated value of $\Delta Q$. The value of $B$ was constant for $Q > 0.3$ \AA$^{-1}$, where the elastic component almost consists of incoherent elastic scattering. The energy resolution was obtained to be $\Delta\omega$ = 2.0 meV, i.e., $\Delta\omega/E_i$ = 2\%.
\begin{figure}
\includegraphics[width=8cm]{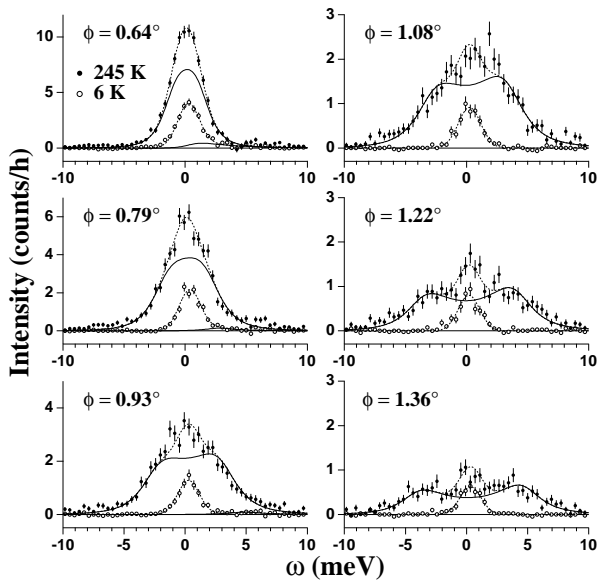}
\caption{\label{fig:epsart} Energy spectra for La$_{0.8}$Sr$_{0.2}$MnO$_3$ at $T$ = 6 and 245 K observed on the scan loci at $\phi$. The dashed lines are fitted curves to the observed spectra, and the solid lines are spin wave components (see text).}
\end{figure}

\begin{figure}
\includegraphics[width=8cm]{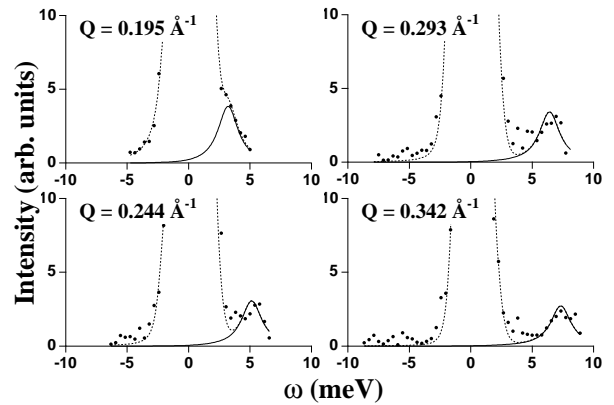}
\caption{\label{fig:epsart} Energy spectra for SrRuO$_3$ at $T = 7$ K obtained by a constant $Q$ cut, which are integrated intensities within the range $Q \pm 0.024$ \AA$^{-1}$ of the intensity map in the $(Q,\omega)$ space. The dashed lines are fitted curves to the observed spectra and the solid lines are spin wave components (see text).}
\end{figure}

Because every position in any PSD can be converted into an appropriate $\phi$, an intensity map in the $(Q,\omega)$ space can be obtained by averaging the intensities on a ring, which require to all have the same $Q$, by using all the areas of the PSD array. Figure 3 presents the energy spectra spanned on the $(Q,\omega)$  space for SrRuO$_3$ at $T$ = 7 K. It corresponds to constant $Q$ cuts, which were integrated intensities within the range $Q \pm 0.024$ \AA$^{-1}$. Each spectrum was well fitted to eq. (4) for $\omega > 0$ meV and the equation in which $n(\omega)+1$ in eq. (4) was replaced by $n(\omega)\exp(\omega/k_BT)$ for $\omega < 0$ meV. The peak positions of the spin wave component were obtained from the parameters $\omega_0$ and $\Gamma$. 
As shown in Fig.3, the peak intensity of the spin wave was nearly constant in $Q$, which was consistent with the powder averaged dynamical structure factor for ferromagnetic spin-waves, because the dispersion relation was approximated with a quadratic form as described below.
Similarly, the peak positions of the spin wave component were obtained from the constant $Q$ cuts for La$_{0.8}$Sr$_{0.2}$MnO$_3$; they were also obtained from the energy spectra on the scan loci at corresponding $\phi$ for SrRuO$_3$. The peak positions of the spin waves obtained from two independent analyses fall onto the single dispersion curve for each scan, as shown in Fig. 4.

\begin{figure}
\includegraphics[width=7.5cm]{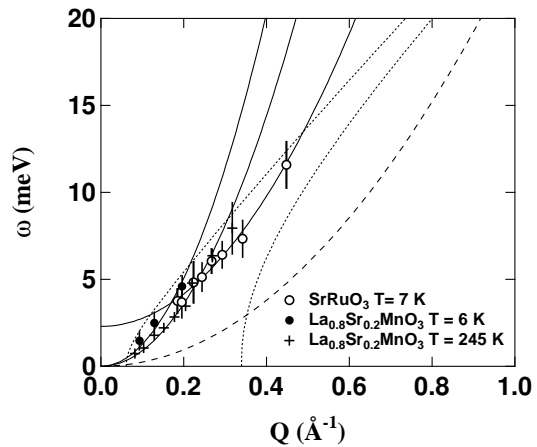}
\caption{\label{fig:epsart} Spin wave dispersion curves of SrRuO$_3$ and La$_{0.8}$Sr$_{0.2}$MnO$_3$. The solid lines are fitted dispersion curves. The dotted lines are the boundaries of the present experiment ($\phi = 0.5^{\circ}$ and $2.8^{\circ}$), and the dashed line is the upper boundary accessible with a spectrometer having the lowest scattering angle of $\phi = 5^{\circ}$ (see Fig.1).}
\end{figure}

The spin wave dispersion curves in La$_{0.8}$Sr$_{0.2}$MnO$_3$ observed at $T$ = 6 and 245 K were well fitted to a quadratic form $\omega = DQ^2$ with $D = 126 \pm 15$ and $91 \pm 5$ meV\AA$^2$, respectively.  In previous works measuring spin waves using single crystals, a dispersion curve without a spin gap with $D$ = 131 meV\AA$^2$ ($T$ = 14 K) and 89 meV\AA$^2$  ($T$ = 250 K) for La$_{0.8}$Sr$_{0.2}$MnO$_3$ was reported \cite{endoh97,moussa07}. At $T$ = 14 K and $Q < 0.3$ \AA$^{-1}$ from an appropriate reciprocal point, the dispersion curve along [001] was well fitted to $\omega = DQ^2$ with $D$ = 131 meV\AA$^2$ and was identical to that along [111] \cite{moussa07}.  Therefore, the spin wave dispersion curves observed in the present powder sample were in good agreement with those in previous studies and confirmed the isotropic dispersion relation for small $Q$. Therefore, the NBS experiment proved effective for using the HRC in the research and development of high performance ferromagnetic materials.

The spin wave dispersion curve of SrRuO$_3$ observed at $T$ = 7 K was well fitted to a quadratic form with an energy gap: $\omega = E_0 + DQ^2$ with $E_0 = 2.3 \pm 0.6$ meV and $D = 47 \pm 8$ meV\AA$^2$. As mentioned above, the crystal lattices for SrRuO$_3$ and La$_{0.8}$Sr$_{0.2}$MnO$_3$ were assumed to have the same simple cubic structure so that the 6 neighboring magnetic ions were distributed along the three principal axes. In the mean field approximation, the Currie temperature is expressed as $T_c = 2zS(S+1)J/3$ with $z$ = 6 being the coordination number. The stiffness constant of spin waves is given by the linear spin wave theory of the ferromagnetic Heisenberg model to be $D = 2zSJa^2$, with lattice constant $a$ being the distance between neighboring magnetic atoms. The $D$ value obtained in SrRuO$_3$ corresponds to $T_c = 151 \pm 26$ K in the mean field approximation, which was in good agreement with the actual Curie temperature ($T_c$ = 165 K). Such a scaling between $D$ and $T_c$ has been commonly observed in metal ferromagnets \cite{endoh06}. A ferromagnet having a high symmetry of crystalline structure usually shows gapless spin-wave excitations due to negligible single ion anisotropy, and, in fact, spin waves in La$_{0.8}$Sr$_{0.2}$MnO$_3$ exhibit a gapless dispersion relation. We clearly observed a large energy gap in the spin wave excitations in SrRuO$_3$, as described herein. This characteristic feature originates from an induced spin anisotropy, possibly from robust spin-orbit coupling acting on the Ru 4d orbital \cite{fang03}. The large anomalous Hall effect, changing the sign in the middle between $T_c$ and the base temperature in the ferromagnetic phase of SrRuO$_3$ was theoretically interpreted as a strong spin-orbit interaction \cite{fang03}. This theory also predicted a magnetic monopole scenario, which must be proven by future experiments. Spin wave excitations in the present low $Q$ region reveal a robust spin-orbit coupling effect in this crystal for the first time. 

In conclusion, we developed an NBS method using a pulsed spallation neutron source and demonstrated its feasibility by observing ferromagnetic spin waves in SrRuO$_3$ and La$_{0.8}$Sr$_{0.2}$MnO$_3$ powders. It should be emphasized again here that the short pulse and high intensity of the neutron source in a sub-eV region, as well as the performance of the HRC, with its high energy resolution and low background, made the NBS method possible. It showed that powdered samples could be used for neutron spectroscopy. This demonstration also shows a well-defined quadratic $Q$ dependence in the spin wave dispersion curve and a large energy gap in the SrRuO$_3$ ferromagnet, which demonstrates the importance of the spin-orbit coupling effect.

This neutron scattering experiment was approved by the Neutron Scattering Program Advisory Committee of the Institute of Materials Structure Science, High Energy Accelerator Research Organization (Proposal Nos. 2011S01 and 2012S01). S. I. and T. Y. were partially supported by a Grant-in-Aid for Scientific Research (C), and Y. E. was supported by a Grant-in-Aid for Challenging Exploratory Research, both were from the Japanese Ministry of Education, Culture, Sports, Science and Technology. This work was also partly supported by the Funding Program for World-Leading Innovative R\&D on Science and Technology (FIRST Program) from the Japan Society for the Promotion of Science. 


\end{document}